\documentclass[iop,apj]{emulateapj}
\usepackage{amsmath}
\usepackage{mathrsfs}
\usepackage[varg]{txfonts}
\usepackage{braket}
\usepackage{cancel}
\usepackage{bm}

\usepackage{CJK}
\usepackage[utf8]{inputenc}

\newtheorem{theorem}{Theorem}
\newtheorem{lemma}[theorem]{Lemma}
\newtheorem{corol}[theorem]{Corollary}

\renewcommand\le\leqslant
\renewcommand\ge\geqslant

\newcommand\ham{\mathcal{H}}
\newcommand\df{\mathcal{F}}
\newcommand\dst{\mathcal{D}}
\newcommand\iom{\mathcal{I}}

\newcommand\kname{\CJKfamily{mj}안진혁}

\begin{document}
\begin{CJK*}{UTF8}{}

\journalinfo{to appear in
\it the Astrophysical Journal \rm (2015 November 12)}

\title{What can the alignments of the velocity moments tell us
\\about the nature of the potential?}
\shorttitle{Velocity Moments of Separable Potentials}

\author{J.~An (\kname)\altaffilmark{1} and N.~W.~Evans\altaffilmark{2}}
\altaffiltext{1}
{National Astronomical Observatories, Chinese Academy of Sciences,
A20 Datun Rd., Chaoyang Dist., Beijing 100012, China}
\altaffiltext{2}
{Institute of Astronomy, University of Cambridge,
Madingley Road, Cambridge CB3 0HA, UK}
\shortauthors{An \& Evans}
\slugcomment{accepted 2015 November 13 (revised version);
submitted 2015 September 29 (original version)}

\begin{abstract}\noindent
We prove that, if the time-independent
distribution function $\df(\bm v;\bm r)$
of a steady-state stellar system is symmetric under velocity inversion
such that $\df(-v_1,v_2,v_3;\bm r)=\df(v_1,v_2,v_3;\bm r)$ and the same for
$v_2$ and $v_3$, where $(v_1,v_2,v_3)$ is the velocity component
projected onto an orthogonal frame, then the potential within which
the system is in equilibrium must be separable (i.e.\ the St\"ackel
potential). Furthermore, we find that the Jeans equations imply that,
if all mixed second moments of the velocity vanish; that is,
$\langle v_iv_j\rangle=0$ for any $i\ne j$, in some St\"ackel coordinate
system and the only non-vanishing fourth moments in the same coordinate
are those in the form of $\langle v_i^4\rangle$ or $\langle v_i^2v_j^2\rangle$,
then the potential must be separable in the same
coordinates. Finally we also show that all second and fourth velocity
moments of tracers with an odd power to the radial component $v_r$
being zero is a sufficient condition to guarantee the potential to be
of the form $\Phi=f(r)+r^{-2}g(\theta,\phi)$.
\end{abstract}

\section{Introduction}

With the advent of large data sets for stellar kinematics in the solar
neighborhood, there are growing evidences to suggest that the velocity
ellipsoids constructed from the local halo stars are aligned along the
coordinate frame directions of the spherical polar coordinate centered
at the Galactic center to a good
approximation \citep[e.g.,][]{SEA,Bo10,Ki15}. \citet{SEA} have
claimed that this alignment implies the sphericity of the underlying
potential due to the Galactic dark matter halo. Their argument is
based on the theorem that, if all orbits in the given potential
respect an integral of motion that is independent of the sign for the
radial component of the velocity, then the radial coordinate can always be
separated off in the Hamilton--Jacobi equation for the system. Thanks
to the Jeans theorem, this indicates that steady-state
populations with a distribution symmetric under the parity of the
radial motion are allowed only if the gravitational potential is
spherical or in the form of the one due to a pure dipole, while the
velocity ellipsoids resulting from such populations must be aligned
radially in the direction of the spherical coordinate frames.

Nevertheless this reasoning is incomplete because there may exist
distribution functions that fail the symmetry condition, but still
produce velocity ellipsoids aligned radially. Since the velocity
ellipsoid is defined for any distribution irrespective of its
symmetry, it is always possible to find such a distribution
\emph{locally}. However, the answer to the question as to whether it
is possible to construct a global steady-state distribution function,
that does not possess symmetry under reversal of the radial velocity,
but which none the less has radially aligned velocity ellipsoids
everywhere, is still unclear at the moment. Note that \citet{BM11}
have provided a three-integral distribution function in a highly
flattened axisymmetric potential with the velocity ellipsoids at some
high-latitude locations aligned radially, albeit not globally. In
fact, they have argued that the explicit connection between the
behavior of the velocity ellipsoids and the shape of the potential can
only be drawn under the prior assumption of the St\"ackel potential, but
not with an arbitrary potential.

We then would like to ask what are the observational constraints to
guarantee the underlying potential to be separable in the given
coordinate. Classically, a sufficient condition for the St\"ackel
potential is that the distribution function is given by a function of
a quadratic polynomial of velocities (other than combinations of the
energy and the square of the angular momentum) \citep[c.f.][]{Ed15,Ch39}.
Very recently, \citet{Ev15} have also shown
that there actually exists a weaker sufficient condition on the
distribution for separable potentials; namely, if the even part of
the distribution is symmetric under each separate parity transform
of a single momentum component in spherical, cylindrical, or
spheroidal/ellipsoidal coordinate, then the potential must be in the
separable form in the corresponding coordinate. Although this does
not settle the original question regarding whether the alignment of
the velocity ellipsoids can by itself imply the separability of the
potential, we conjecture that ``alignments'' of even velocity moments
in \emph{every} order actually can. This idea will be formalized rigorously
and proven in this paper.

Extending this, we also seek possibility of relaxing the
requirement for every order to some finite subsets of the velocity
moments. An obvious line of approach would be using the Jeans
equations, which directly relate velocity moments and their
spatial gradients to the underlying potential. While there have been
many investigations based on the Jeans equation to find a model with
aligned velocity ellipsoids in an arbitrary potential, most, if not all,
of these studies have only considered the behavior of the velocity
dispersions (i.e.\ the second moments). By contrast, the symmetric
distribution considered by \citet{Ev15} actually generates a
specifically constrained set of velocity moments of higher order, too.
Inspired by this, we consider what constraints on the potential
may be deduced if additional conditions on the alignments of higher
order moments are imposed. In the end, we discover that the alignments
of the fourth moments as well as the second moments are actually
a sufficient assumption to deduce the separability of the potential.

In the following section (\S~\ref{sec:pre}),
we first review the principal concepts
related to the St\"ackel coordinates and the separable potential.
Compared to the standard approach typically found in the astrophysical
literature \citep[see e.g.,][]{dZ85,dZL85}, the point of view here is
slightly more abstract and somewhat more formal, which is more
suitable for our purpose and also affords us more general
conclusions. For sake of the self-containedness, we include
more materials than what is absolutely necessary. In \S~\ref{sec:dss},
we then generalize the result of \citet{Ev15} and provide its more
formal proof. In addition, we also introduce a precise statement
concerning the alignment of the higher order velocity moments in terms
of vanishing crossing terms. In \S~\ref{sec:je}, we then derive the
Jeans equations in every order from the moment integrals on the
collisionless Boltzmann equation, which are the basis of the proof
found in the subsequent sections. The next section (\S~\ref{sec:sfmsp})
provides the proof of the primary result of this paper, Theorem \ref{theo-ms};
that is, the alignments of the second and fourth moment in the St\"ackel
coordinate implies the separability of the potential in the same
coordinate. In \S~\ref{sec:psp3d}, we shift our focus to specific
3-dimensional cases and find that the translational, rotational, or
spherical symmetry of the potential may be inferred from only subsets
of the alignment requirement for the second and fourth moments.

\section{Preliminaries}\label{sec:pre}
\subsection{the St\"ackel coordinate}

In the astrophysical literature, the St\"ackel coordinates are
usually considered as synonymous to the confocal ellipsoidal
coordinates (including their degenerate limits). Although this approach
is not necessarily incorrect, it is still unsatisfactory because it
does not inform us about their defining characteristics. Instead, we
consider a pedagogical definition of the St\"ackel coordinate; namely,
the orthogonal curvilinear coordinate in which the Hamilton--Jacobi
equation (HJE) for the geodesic (i.e.\ force-free) motion is soluble
through additive separation of variables.

Provided that the Hamiltonian $\ham(q^1,\dotsc,q^n;p_1,\dotsc,p_n)$
does not explicitly depend on the time, the HJE is reducible to the
partial differential equation on the Hamilton characteristic function,
$W(q^1,\dotsc,q^n)$; that is, $\ham(q^1,\dotsc,q^n;W_{,1},\dotsc,W_{,n})=E$,
where $W_{,i}\equiv\partial W/\partial q^i$ and $E$ is a constant.
The reduced HJE then implies
$(\partial/\partial q^k)\ham(q^1,\dotsc,q^n;W_{,1},\dotsc,W_{,n})=0$:
that is,
\begin{equation}\textstyle
\partial_k\ham\rvert_{p_i=W_{,i}}
+\sum_{j=1}^nW_{,jk}\partial^j\ham\rvert_{p_i=W_{,i}}=0
\end{equation}
for any $k$. Here $\partial_k\ham
\equiv\partial\ham(q^1,\dotsc,q^n;p_1,\dotsc,p_n)/\partial q^k$ and
$\partial^j\ham\equiv
\partial\ham(q^1,\dotsc,q^n;p_1,\dotsc,p_n)/\partial p_j$.
The Hamilton--Jacobi method of integrating the equation of
motions seeks a complete solution of this equation such that
$W=\sum_{i=1}^nw_i(q^i)$ (so $W_{,ij}=0$ for all $i\ne j$). Existence
of a complete solution is equivalent to existence of the solution set
$\set{p_1,\dotsc,p_n}$ for the overdetermined system of the partial
differential equations,
\begin{equation}\label{eq:pde}
\frac{\partial p_j}{\partial q^j}
=-\frac{\partial_j\ham}{\partial^j\ham}\,;\quad
\frac{\partial p_j}{\partial q^k}=0
\quad(j\ne k).
\end{equation}
In order for this system to be integrable,
the compatibility condition; that is,
$(\partial/\partial q_k)(\partial p_j/\partial q_j)
=(\partial/\partial q_j)(\partial p_j/\partial q_k)=0$ for all $j\ne k$
needs to be satisfied. This then results in
\begin{multline}\label{eq:LCc}\textstyle
(\partial^j\ham)(\partial^k\ham)(\partial_k\partial_j\ham)
+(\partial_j\ham)(\partial_k\ham)(\partial^k\partial^j\ham)
\\=(\partial^j\ham)(\partial_k\ham)(\partial^k\partial_j\ham)
+(\partial_j\ham)(\partial^k\ham)(\partial_k\partial^j\ham)
\end{multline}
for any $j\ne k$, which is known as
the Levi-Civita separability condition after \citet{LC04}.

In an orthogonal curvilinear coordinate $(q^1,\dotsc,q^n)$
with the scale factors $\set{h_1,\dotsc,h_n}$, the Hamiltonian of
the geodesic motion is given by $\ham=\sum_{i=1}^np_i^2/(2h_i^2)$.
The Levi-Civita condition on this Hamiltonian reduces to
\begin{equation}\label{eq:stk}\textstyle
\sum_{i=1}^np_i^2\dst_{jk}(h_i^{-2})=0
\quad(j\ne k)
\end{equation}
where $\dst_{jk}(f)$ is the differential operator acting
on a function $f(q^1,\dotsc,q^n)$, defined to be
\begin{equation}\label{eq:ddef}
\dst_{jk}(f)\equiv\frac{\partial^2f}{\partial q^j\partial q^k}
+\frac{\partial\ln h_k^2}{\partial q^j}\frac{\partial f}{\partial q^k}
+\frac{\partial\ln h_j^2}{\partial q^k}\frac{\partial f}{\partial q^j}.
\end{equation}
Here note that $\dst_{jk}=\dst_{kj}$. Hence
the necessary and sufficient condition for the HJE of
the geodesic Hamiltonian in the chosen orthogonal coordinate
to be soluble through separation of variable is
$\dst_{jk}(h_i^{-2})=0$ for all triplets of indices $(i,j,k)$ with
$j\ne k$, which is referred to as the St\"ackel coordinate condition
after \citet{St91,St93}. That is to say, St\"ackel coordinates are
any orthogonal curvilinear coordinates whose scale factors satisfy the
St\"ackel coordinate condition.

Like the Levi-Civita condition,
the St\"ackel coordinate condition is also understood to be
the integrability condition for existence of the solution set
to a system of differential equations. In particular, suppose that
there exists a set of $n$ independent functions
$\set{u_1(q^1),\dotsc,u_n(q^n)}$ such that
$U(q^1,\dotsc,q^n)=\sum_{i=1}^nu_i(q^i)/h_i^2$ is constant.
Then $\bm\nabla U=0$, or
\begin{equation}
\frac{\partial U}{\partial q^k}
=\sum_{i=1}^n\frac{\partial h_i^{-2}}{\partial q^k}u_i(q^i)
+\frac{u_k'(q^k)}{h_k^2}=0,
\end{equation}
for any $k$.
Thus the set of functions $u_i$ must be the solution of
\begin{equation}
\frac{\partial u_j}{\partial q^k}
=-\delta^k_jh_k^2\sum_{i=1}^n\frac{\partial h_i^{-2}}{\partial q^k}u_i,
\end{equation}
where $\delta_j^k$ is the Kronecker delta. The integrability condition
on this set of partial differential equations then results in
\begin{equation}
\frac\partial{\partial q^j}\cancel{\frac{\partial u_j}{\partial q^k}}
-\frac\partial{\partial q^k}\frac{\partial u_j}{\partial q^j}
=h_j^2\sum_{i=1}^n\dst_{jk}(h_i^{-2})\,u_i=0
\quad(j\ne k).
\end{equation}
In other words, the condition that $\dst_{jk}(h_i^{-2})=0$
for all $j\ne k$ and any $i$
implies existence of the set of functions $u_i(q^i)$ such that
$\sum_{i=1}^nu_i/h_i^2$ is constant. Moreover,
the Frobenius theorem further indicates that there actually exist $n$
such linearly-independent solution sets $\set{u^j_1(q^1),\dotsc,u^j_n(q^n)}$
where $j\in\set{1,\dotsc,n}$. Hence, if $h_i$'s are the scale factors of
the St\"ackel coordinate, there exists an invertible
($n\times n$)-matrix of functions $[S^j_i(q^i)]$ that satisfy constraints:
\begin{equation}\label{eq:stkm}
\sum_{i=1}^n\frac{S_i^j(q^i)}{h_i^2}
=\begin{cases}1&(j=1)\\0&(j=2,\dotsc,n)\end{cases}.
\end{equation}
This is equivalent to insisting that $h_i^{-2}=C_i^1/\lvert S\rvert$,
where $C_i^j$ is the co-factor of the matrix $[S^j_i(q^i)]$ and
$\lvert S\rvert=\det\{S^j_i(q^i)\}=\sum_{i=1}^nS_i^jC_i^j$ is its determinant
(known as the St\"ackel determinant). Existence of such invertible
matrices of functions $[S^j_i(q^i)]$ may be considered as an alternative
definition of the St\"ackel coordinate \citep[c.f.][]{Go80}, which is
closer to St\"ackel's original approach.

The St\"ackel coordinate condition is the system of partial
differential equations on the scale factors of an orthogonal coordinate.
It is fairly straightforward to demonstrate that the scale factors of
the confocal ellipsoidal coordinate as well as all of its degenerate limits
satisfy the St\"ackel coordinate condition.
On the other hand, the differential equation system due to
the St\"ackel coordinate condition can in principle be soluble
to obtain the general expression (including some arbitrary functions)
for the scale factors of the St\"ackel coordinates.
In the flat Euclidean space, the general solution actually results
in the scale factors of the confocal ellipsoidal coordinate,
up to arbitrary scaling functions \citep{LC04}. The same result
was also found by \citet{Ed15} and \citet{Ly62}, although
their respective assumptions upon which the derivation
of the differential equations equivalent to the St\"ackel coordinate
condition is based are distinct from the consideration here.

\subsection{The separable or St\"ackel potentials}

Next let us consider the condition for the HJE of a natural dynamical
system with the potential $\Phi(q^1,\dotsc,q^n)$ to be soluble via
separation of variables. In an orthogonal coordinate, the Hamiltonian
of a natural system is $\ham=\sum_{i=1}^np_i^2/(2h_i^2)+\Phi$, and the
Levi-Civita condition simplifies to
\begin{equation}\textstyle
\frac12\sum_{i=1}^np_i^2\dst_{jk}(h_i^{-2})
+\dst_{jk}(\Phi)=0
\quad(j\ne k).
\end{equation}
Hence the corresponding HJE admits a complete integral
if the chosen orthogonal coordinate is the St\"ackel coordinate
and the potential is the solution of the differential equation
$\dst_{jk}(\Phi)=0$ for all $j\ne k$ with $\dst_{jk}$
given by equation (\ref{eq:ddef}).

With $\Phi=\sum_{i=1}^nf_i(q^i)/h_i^2$, where $f_i(q^i)$ is an
arbitrary function of the coordinate component $q^i$ alone and $h_i$'s
are the scale factors of the St\"ackel coordinate, it is straightforward
to show $\dst_{jk}(\Phi)=0$ for all $j\ne k$. If the specific expression of
the scale factors are given, the opposite implication is also shown to
hold by solving the differential equation. For general cases however,
one needs to find the system of partial differential equations,
whose integrability condition leads to $\dst_{jk}(\Phi)=0$. In
particular, if we assume existence of the set of functions
$\set{f_1(q^1),\dotsc,f_n(q^n)}$ such that
$\Phi=\sum_{i=1}^nf_i(q^i)/h_i^2$, then
\begin{equation}
\frac{\partial\Phi}{\partial q^j}
=\frac{f_j'(q^j)}{h_j^2}
+\sum_{i=1}^n\frac{\partial h_i^{-2}}{\partial q^j}f_i(q^i),
\end{equation}
and so $\set{f_1,\dotsc,f_n}$ must be the solution set of the system
\begin{equation}
\frac{\partial f_k}{\partial q^j}
=\delta^j_kh_j^2\left(\frac{\partial\Phi}{\partial q^j}
-\sum_{i=1}^n\frac{\partial h_i^{-2}}{\partial q^j}f_i\right).
\end{equation}
According to the Frobenius theorem, the necessary and sufficient condition
for the solution to exist is the compatibility condition
$(\partial/\partial q^j)(\partial f_i/\partial q^k)
=(\partial/\partial q^k)(\partial f_i/\partial q^j)$ for any $i,j,k$
to hold. Here the only non-trivial conditions among them are
\begin{equation}
\frac\partial{\partial q^j}\left(\frac{\partial f_k}{\partial q^k}\right)
=h_k^2\left(\dst_{jk}(\Phi)
-\sum_{i=1}^n\dst_{jk}(h_i^{-2})\,f_i\right)=0
\end{equation}
for all $j\ne k$.
The condition $\dst_{jk}(\Phi)=0$ is thus the necessary condition
for existence of the solution set $\set{f_1,\dotsc,f_n}$,
while the same condition is also sufficient for (local) existence of
such solution sets with the scale factors of the St\"ackel coordinate
satisfying $\dst_{jk}(h_i^{-2})=0$ for $j\ne k$. In other words,
\begin{equation}\label{eq:sepp}
\Phi(q^1,\dotsc,q^n)=\sum_{i=1}^n\frac{f_i(q^i)}{h_i^2}
\end{equation}
is the general solution of $\dst_{jk}(\Phi)=0$ for all $j\ne k$, given
$\dst_{jk}(h_i^{-2})=0$. Henceforth, we shall refer the potential in
the form of equation (\ref{eq:sepp}) to be separable in the particular
St\"ackel coordinate, whereas the potential shall be referred to as
the St\"ackel potential if there exists a St\"ackel coordinate in
which the potential is expressible as in equation (\ref{eq:sepp}).

By definition, the HJE of the natural dynamical system with
a St\"ackel potential is soluble through separation of variables in the
St\"ackel coordinate in which the potential is separable. Less
abstractly, the St\"ackel potential admits a set of $n$ independent
integrals of motion. In particular, let
\begin{equation}\label{eq:int}\textstyle
\alpha_j=\sum_{i=1}^n
[p_i^2+2f_i(q^i)]\,T^i_j(q^1,\dotsc,q^n),
\end{equation}
where $(T^i_j)$ is the inverse matrix of $(S_i^j)$
in equation (\ref{eq:stkm}) for the chosen coordinate
(i.e.\ $T^i_j=C^j_i/\lvert S\rvert$).
Here $T^i_1=C^1_i/\lvert S\rvert=h_i^{-2}$ and so $\alpha_1=2\ham$.
Next consider the Poisson brackets
\begin{equation}
\left\lbrace\alpha_j,\alpha_k\right\rbrace
=\sum_{i,\ell=1}^n2p_i(p_\ell^2+2f_\ell)
\left(T^i_k\frac{\partial T^\ell_j}{\partial q^i}
-T^i_j\frac{\partial T^\ell_k}{\partial q^i}\right).
\end{equation}
However, since $\sum_{\ell=1}^nS_i^\ell T_\ell^j=\delta_i^j$
and $S_i^\ell=S_i^\ell(q^i)$, we find
\begin{equation}
\sum_{\ell=1}^nS_i^\ell\frac{\partial T_\ell^j}{\partial q^k}
=\begin{cases}
R_k^j&(i=k)\\0&(i\ne k)
\end{cases};\quad
R_k^j=-\sum_{\ell=1}^n\frac{dS_k^\ell}{dq^k}T_\ell^j,
\end{equation}
which further implies that
\begin{equation}
\frac{\partial T^\ell_j}{\partial q^i}
=\sum_{m,k=1}^nT^m_jS_m^k\frac{\partial T^\ell_k}{\partial q^i}
=T^i_jR_i^\ell,
\end{equation}
and thus $\lbrace\alpha_j,\alpha_k\rbrace=0$ for any $j,k$. In other words,
$\alpha_i$'s are all functionally-independent -- thanks to $(T^i_j)$
being invertible -- integrals of motion (note $\alpha_1=2\ham$) that
are in involution and so all orbits within the St\"ackel potential are
Liouville-integrable (i.e.\ all bounded orbits are quasi-periodic).

We note that every integral of motion $\alpha_j$ is a linear function
of $p_i^2$'s. What is more interesting is the converse: namely,
\begin{theorem}\label{theo-qd}
If the natural dynamical system admits an integral of motion
expressible in an orthogonal coordinate as
$\mathscr I=\sum_{i=1}^n\zeta_iv_i^2+\Xi$, where $\zeta_i$'s and $\Xi$ are
smooth functions of positions and all $\zeta_i$'s are distinct,
then the coordinate must be a St\"ackel coordinate and
the potential is separable in the same coordinate.
\end{theorem}
{\it Proof:}
First let the Hamiltonian be $\ham=\sum_{i=1}^np_i^2/(2h_i^2)+\Phi$.
Then $v_i^2=p_i^2/h_i^2$ and so $\mathscr I$ is an integral of motion if
\begin{equation}
\left\lbrace\mathscr I,\ham\right\rbrace=\sum_{j=1}^n
\left(\sum_{i=1}^n\mathscr A_{ij}p_i^2+\mathscr B_j\right)\frac{p_j}{h_j^2},
\end{equation}
identically vanishes, where
\begin{equation}
\mathscr A_{ij}\equiv
\frac\partial{\partial q^j}\left(\frac{\zeta_i}{h_i^2}\right)
-\zeta_j\frac{\partial h_i^{-2}}{\partial q^j};\quad
\mathscr B_j\equiv\frac{\partial\Xi}{\partial q^j}
-2\zeta_j\frac{\partial\Phi}{\partial q^j}.
\end{equation}
This requires all $\mathscr A_{ij}=0$ and $\mathscr B_j=0$: that is,
for any $i,j$,
\begin{equation}\label{eq:edd}
\frac{\partial\zeta_i}{\partial q^j}
=(\zeta_j-\zeta_i)\frac{\partial\ln h_i^{-2}}{\partial q^j};\quad
\frac{\partial\Xi}{\partial q^j}=2\zeta_j\frac{\partial\Phi}{\partial q^j}.
\end{equation}
Here $\mathscr A_{ij}=0$ results in a system of differential equations
on $\zeta_i$'s, and so in order for the solution to exist, the compatibility
condition should again be satisfied: namely,
\begin{equation}
\frac\partial{\partial q^k}\frac{\partial\zeta_i}{\partial q^j}
-\frac\partial{\partial q^j}\frac{\partial\zeta_i}{\partial q^k}
=(\zeta_j-\zeta_k)\,h_i^2\dst_{jk}(h_i^{-2})=0.
\end{equation}
Provided that $\zeta_j\ne\zeta_k$ for $j\ne k$, the St\"ackel
coordinate condition is therefore indeed necessary for
$\lbrace\mathscr I,\ham\rbrace=0$. Similarly the
compatibility condition on $\Xi$
results\footnote{This may be derived alternatively as follows:
 i.e.\ existence of a function $\Xi(q^1,\dotsc,q^n)$ indicates
 that the vector field $\bm\nabla\Xi=\sum_i\bm e^i\partial_i\Xi
 =2\sum_i\bm e^i\zeta_i\partial_i\Phi$ (where $\bm e^i\equiv\bm\nabla q^i$
 is the co-frame vector), must be curl-free. Since $\partial_i\zeta_i=0$
 and $\bm\nabla\wedge\bm e^i=0$, we have
 $\bm\nabla\wedge(\bm e^i\zeta_i\partial_i\Phi)
 =[\bm\nabla\partial_i(\zeta_i\Phi)]\wedge\bm e^i$
 and so the curl-free condition is explicitly given by
 $\bm\nabla\wedge\bm\nabla\Xi=2\sum_{i,j}\bm e^j\wedge\bm e^i
 \partial_j\partial_i(\zeta_i\Phi)
 =\sum_{i,j}\bm e^j\wedge\bm e^i
 \partial_j\partial_i[(\zeta_i-\zeta_j)\Phi]=0$,
 which is equivalent to eq.~(\ref{eq:pzc}) given
 $\partial_i\partial_j[(\zeta_i-\zeta_j)\Phi]
 =(\zeta_i-\zeta_j)\dst_{ij}(\Phi)$.}
in
\begin{equation}\label{eq:pzc}
\frac\partial{\partial q^j}\frac{\partial\Xi}{\partial q^i}
-\frac\partial{\partial q^i}\frac{\partial\Xi}{\partial q^j}
=(\zeta_i-\zeta_j)\,\dst_{ij}(\Phi)=0,
\end{equation}
and so this indicates that $\dst_{ij}(\Phi)=0$ for all $i\ne j$,
assuming $\zeta_i\ne\zeta_j$ for any $i\ne j$. {\sc qed.}

This theorem was implicit in \citet{Ed15}, who derived equation
(\ref{eq:edd}) for 3 dimension (his eq.~13) under the so-called
Schwarzschild ellipsoidal hypothesis; that is to say,
$\df\propto\exp(-\mathscr I)$, where $\df$ is the phase-space
distribution function. He then showed that, in the
3-dimensional flat Euclidean space, this implies i) the coordinate
surfaces are confocal quadrics and so the coordinate must be a
confocal ellipsoidal coordinate or one of its degenerate limits and
ii) the potential must be able to be expressible in the form of equation
(\ref{eq:sepp}). Given the \citet{Je15} theorem,
the distribution $\df$ is an integral of motion, and therefore
the ellipsoidal hypothesis implies that $\mathscr I$ is an integral.
In fact, his results were due to $\mathscr I$ being an
integral and do not rely on the assumed form of $\df$. It was not
until \citet{Ly62} that it was explicitly stated that the St\"ackel
potential is implied by existence of an integral of motion in a
specific nature rather than the particular form of the distribution function.

\citet{Ch39} investigated a nominally weaker assumption than that of
\citet{Ed15}; that is, the existence of a distribution $\df(\mathscr I)$
depending on the single integral $\mathscr I$. However, there appears
to be some incompleteness in Chandrasekhar's analysis, as he failed to
identify the St\"ackel potentials as solutions, although
he did find an unusual (albeit somewhat academic) stellar system
with helical symmetry -- see \citet{Ev11} for a historical review.
Technically, the form of the integral $\mathscr I$ considered by
\citet{Ch39} is more relaxed than that of Theorem \ref{theo-qd}, as it
is a quadratic polynomial of the velocity components. Thanks to the
time reversal symmetry of the natural dynamical system, the even and
odd parts of any integral of motion are also independent integrals,
and so his assumption is basically equivalent to existence of an
integral of the form $\mathscr I=q(\bm v)+\Xi$ with $q(\bm v)$ being a
homogeneous degree-two polynomial (i.e.\ a quadratic form) of the
velocities. However, an arbitrary quadratic form (over the reals) can
always be diagonalized\footnote{Thanks to the spectral theorem, this
is achieved through a point-wise orthogonal transform, which defines
an orthonormal frame field over the space via its eigenvectors.
Provided that the frames are differentiable, the Cartan theory
indicates that one can always find a coordinate system such that one
of the vector fields in the frame is normal to one of the coordinate
surfaces.} to bring it into the form considered in Theorem \ref{theo-qd},
although the principal values, $\zeta_i$'s of the integral
are not necessarily all distinct. In other word, Theorem \ref{theo-qd}
actually indicates that existence of any integral that is a quadratic
function of the velocity (excluding some degenerate cases corresponding
to the Hamiltonian or the squares of the momenta) necessarily
implies that the potential is of St\"ackel \citep[see][]{MSVW,Ev90}.

\section{Distributions in the St\"ackel potential}
\label{sec:dss}

\citet{Ev15} have shown that the sufficient condition for the integral
of motion to guarantee the separability of the potential can be weaker
than that of Theorem \ref{theo-qd}. In particular, they found that the
symmetry of the integral under velocity inversion actually
suffices: that is,
\begin{theorem}\label{theo-ev}
Suppose that $(q^1,\dotsc,q^n)$ is an orthogonal coordinate,
and $(p_1,\dotsc,p_n)$ the conjugate set of the momenta.
If the dynamical system governed by the Hamiltonian
$\ham=\sum_{i=1}^np_i^2/(2h_i^2)+\Phi(q^1,\dotsc,q^n)$ observes
an integral of motion of the form
$\iom=\iom(p_1^2,\dotsc,p_n^2;q^1,\dotsc,q^n)$
with $\zeta_j\ne\zeta_k$ for all $j\ne k$, where
$\zeta_i\equiv2h_i^2[\partial\iom/\partial(p_i^2)]$, then
the orthogonal coordinate must be a St\"ackel coordinate and
the potential $\Phi$ is separable in the same coordinate.
\end{theorem}
{\it Proof:}
Since $\iom(p_1^2,\dotsc,p_n^2;q^1,\dotsc,q^n)$ is an integral of motion,
\begin{equation}
\left\lbrace\iom,\ham\right\rbrace
=\sum_{i=1}^np_i\mathscr C_i(p_1^2,\dotsc,p_n^2;q^1,\dotsc,q^n)=0,
\end{equation}
where (note $\partial\iom/\partial p_i=2p_i[\partial\iom/\partial(p_i^2)]$)
\begin{equation}
\mathscr C_i
\equiv\frac1{h_i^2}\frac{\partial\iom}{\partial q^i}
-2\frac{\partial\ham}{\partial q^i}
\frac{\partial\iom}{\partial(p_i^2)}.
\end{equation}
Since both $\iom$ and $\ham$ are invariant under $p_j\to-p_j$ for any $j$,
all $\mathscr C_i$'s are also invariant under the same
transforms. Hence it follows $\sum_{i=1}^np_i\mathscr C_i=0$ that
$\mathscr C_i=0$ for all $i$. Specifically
\begin{equation}
\frac{\partial\iom}{\partial q^i}
=\zeta_i\frac{\partial\ham}{\partial q^i},\quad
\zeta_i\equiv2h_i^2\frac{\partial\iom}{\partial(p_i^2)}.
\end{equation}
Here, we first note that, for any $i,j$
\begin{equation}\begin{split}
\frac{\partial\zeta_i}{\partial q^j}
&=2h_i^2\frac{\partial^2\iom}{\partial q^j\partial(p_i^2)}
+2\frac{\partial h_i^2}{\partial q^j}\frac{\partial\iom}{\partial(p_i^2)}
\\&=2h_i^2\frac\partial{\partial(p_i^2)}
\left(\zeta_j\frac{\partial\ham}{\partial q^j}\right)
+\frac{\zeta_i}{h_i^2}\frac{\partial h_i^2}{\partial q^j}
\\&=4h_i^2h_j^2\iom^{ij}
\frac{\partial\ham}{\partial q^j}
+2h_i^2\zeta_j\frac{\partial^2\ham}{\partial q^j\partial(p_i^2)}
+\zeta_i\frac{\partial\ln h_i^2}{\partial q^j}
\\&=4h_i^2h_j^2\iom^{ij}
\frac{\partial\ham}{\partial q^j}
+(\zeta_i-\zeta_j)\frac{\partial\ln h_i^2}{\partial q^j},
\end{split}\end{equation}
where $\iom^{ij}\equiv\partial^2\iom/[\partial(p_i^2)\partial(p_j^2)]$
(note $\iom^{ij}=\iom^{ji}$).
Next the integrability condition on $\iom$ indicates that
\begin{equation}\begin{split}
\frac\partial{\partial q^i}&\frac{\partial\iom}{\partial q^j}
-\frac\partial{\partial q^j}\frac{\partial\iom}{\partial q^i}
=\frac\partial{\partial q^i}
\left(\zeta_j\frac{\partial\ham}{\partial q^j}\right)
-\frac\partial{\partial q^j}
\left(\zeta_i\frac{\partial\ham}{\partial q^i}\right)
\\&=\frac{\partial\zeta_j}{\partial q^i}\frac{\partial\ham}{\partial q^j}
-\frac{\partial\zeta_i}{\partial q^j}\frac{\partial\ham}{\partial q^i}
+\zeta_j\frac{\partial^2\ham}{\partial q^i\partial q^j}
-\zeta_i\frac{\partial^2\ham}{\partial q^j\partial q^i}
\\&=(\zeta_j-\zeta_i)\,\dst_{ij}(\ham)=0
\quad\text{(for any $i,j$)}
\end{split}\end{equation}
where $\dst_{ij}$ is as defined in equation (\ref{eq:ddef}).
In other words, if there exists an integral $\iom$ with $\zeta_i\ne\zeta_j$
for $i\ne j$, then
\begin{equation}\textstyle
\dst_{ij}(\ham)=\frac12\sum_{k=1}^n\dst_{ij}(h_k^{-2})p_k^2
+\dst_{ij}(\Phi)=0,
\end{equation}
which must holds identically for all $p_k$'s and therefore
$\dst_{ij}(h_k^{-2})=\dst_{ij}(\Phi)=0$ for all $i\ne j$
with $\zeta_i\ne\zeta_j$. {\sc qed.}

The steady-state distribution, which is a solution
of the collisionless Boltzmann equation (CBE), is an integral of motion.
Theorem \ref{theo-ev} thus holds with the integral of motion, $\iom$
replaced by the time-independent distribution function, $\df$.
However the distribution such that $\df=\df(p_1^2,\dotsc,p_n^2)$
must be even, $\df(-\bm v)=\df(\bm v)$; that is,
$\df(-p_1,\dotsc,-p_n)=\df(p_1,\dotsc,p_n)$, which is
unnecessarily restrictive, for no bulk streaming motion is allowed.
None the less, with the Hamiltonian invariant under the time reversal,
the equation of motion and all the resulting orbits are also symmetric
under the time reversal. Consequently, if $\df(\bm v)$ is a solution
of the CBE, then $\df(-\bm v)$ must be an integral of motion too.
From these, we find
\begin{corol}\label{corol-df}
Suppose that $(q^1,\dotsc,q^n)$ is an orthogonal coordinate.
If the even part of the steady-state distribution function
$\df^+\equiv\frac12[\df(\bm v)+\df(-\bm v)]$
is symmetric under the sign reversal of the single
conjugate momentum component $p_i$ for each coordinate component $q^i$
-- i.e.\ $\df^+(p_1,\dotsc,-p_i,\dotsc,p_n)=\df^+(p_1,\dotsc,p_i,\dotsc,p_n)$
for each $i$
-- and the principal axes of the velocity second moment tensor
(which must be diagonalized along the frame of the given orthogonal
coordinate) are all distinct (i.e.\ $\langle v_i^2\rangle\ne\langle
v_j^2\rangle$ for all $i\ne j$), then the orthogonal coordinate must
be a St\"ackel coordinate and the potential is separable in the same
coordinate.
\end{corol}
Here, we have replaced the non-degeneracy condition on the integral,
$h_i^2[\partial\iom/\partial(p_i^2)]\ne
h_j^2[\partial\iom/\partial(p_j^2)]$ for all $i\ne j$ by the velocity
second moment tensor with all distinct principal axes.
This is allowed because the general solution of
$h_i^2[\partial\df^+/\partial(p_i^2)]=h_j^2[\partial\df^+/\partial(p_j^2)]$
for $i\ne j$ is $\df^+=\df^+(v_{ij}^2)$ where
$v_{ij}^2=(p_i/h_i)^2+(p_j/h_j)^2=v_i^2+v_j^2$. In other words,
the dependences of $\df^+$ on $(p_i,p_j)$ are only through $v_{ij}^2$
(i.e.\ $\df^+$ becomes isotropic within $v_i$-$v_j$ plane),
which implies $\langle v_i^2\rangle=\langle v_j^2\rangle$.

However, construction of the full distribution is challenging.
Instead the usual constraints on the distribution are typically given
as the set of velocity moments in the orthogonal coordinate: namely
(here $\varrho\equiv\int\!d^n\!\bm v\,\df$ is the local density)
\begin{equation}\begin{split}
\varrho\left\langle\prod_{i=1}^nv_i^{m_i}\right\rangle
&=\int\!d^n\!\bm v\left(\prod_{i=1}^nv_i^{m_i}\right)\df
\\&=\idotsint\frac{dp_1\dotsm dp_n}{\prod_{i=1}^nh_i}
\frac{\prod_{i=1}^np_i^{m_i}}{\prod_{i=1}^nh_i^{m_i}}\,\df.
\end{split}\end{equation}
Note that $\df(\bm v)\to\df(-\bm v)$ results in
$\int\!d^n\!\bm v\,\df(-\bm v)(\prod_iv_i^{m_i})
=\int\!d^n\!\bm v\,\df(\bm v)[\prod_i(-v_i)^{m_i}]
=(-1)^{\sum_im_i}\varrho\langle\prod_iv_i^{m_i}\rangle$, and so
\begin{equation}
\int\!d^n\bm v\left(\prod_{i=1}^nv_i^{m_i}\right)\df^+
=\begin{cases}
\varrho\left\langle\prod_iv_i^{m_i}\right\rangle&
\text{if $\sum_im_i$ is even}
\\0&\text{if $\sum_im_i$ is odd}
\end{cases}.
\end{equation}
If $\df^+(p_1,\dotsc,-p_i,\dotsc,p_n)=\df^+(p_1,\dotsc,p_i,\dotsc,p_n)$,
we also find $\langle v_1^{m_1}\dotsm v_i^{m_i}\dotsm v_n^{m_n}\rangle=
(-1)^{m_i}\langle v_1^{m_1}\dotsm v_i^{m_i}\dotsm v_n^{m_n}\rangle$
for even $\sum_{j=1}^nm_j$.
Thus, the only non-vanishing even velocity moments resulting from
the distribution satisfying the condition of Corollary \ref{corol-df}
are in the form of $\langle\prod_{i=1}^nv_i^{2m_i}\rangle$
-- specifically $\langle v_i^2\rangle$ for the second moments, $\langle
v_i^4\rangle$ and $\langle v_i^2v_j^2\rangle$ for the fourth moments
and so on. In fact, the converse also holds:
\begin{lemma}\label{lem}
Suppose that $\df(p_1,\dotsc,p_n)$ is a phase-space distribution
with $(p_1,\dotsc,p_n)$ the conjugate momentum set
of an orthogonal coordinate $(q^1,\dotsc,q^n)$. Then
$\df(p_1,\dotsc,p_n)=\df(-p_1,\dotsc,p_n)$ if and only if
$\langle p_1^{m_1}\dotsc p_n^{m_n}\rangle=0$ for any odd integer $m_1$,
while $\df^+(p_1,\dotsc,p_n)=\df^+(-p_1,\dotsc,p_n)$
if and only if $\langle p_1^{m_1}\dotsc p_n^{m_n}\rangle=0$ for
all even moments with odd $m_1$.
\end{lemma}
\begin{corol}\label{corol-vm}
If the only non-vanishing even velocity moments of
the steady-state tracers in an orthogonal coordinate are of the form
$\langle\prod_{i=1}^nv_i^{2m_i}\rangle$ and the second moments are all
distinct, then the orthogonal coordinate must be a St\"ackel
coordinate and the potential is separable in the same coordinate.
\end{corol}

The remaining ``if''-part of Lemma \ref{lem} may be proven
utilizing the characteristic function $\varphi$; that is,
consider the Fourier transform of the distribution
(here $\bm p\bm\cdot\bm k\equiv\sum_{\ell=1}^np_\ell k^\ell$),
\begin{equation}\textstyle
\varphi(k^1,\dotsc,k^n)\equiv\idotsint dp_1\dotsm dp_n\,
e^{{\rm i}\bm p\bm\cdot\bm k}\,\df(p_1,\dotsc,p_n).
\end{equation}
The partial derivatives of $\varphi$ with respect to $k^j$'s
result in
\begin{multline}
\varphi^{(m_1,\dotsc,m_n)}(k^1,\dotsc,k^n)\equiv
\left[\prod_{j=1}^n\left(\frac\partial{\partial k^j}\right)^{m_j}\right]\varphi
\\={\rm i}^m\!\idotsint dp_1\dotsm dp_n\,
e^{{\rm i}\bm p\bm\cdot\bm k}\left(\prod_{j=1}^np_j^{m_j}\right)\df.
\end{multline}
where $m=\sum_{j=1}^nm_j$.
Evaluating at $\bm k=0$, this results in
\begin{equation}\textstyle
\varphi^{(m_1,\dotsc,m_n)}(0)
={\rm i}^m\,(\prod_jh_j^{m_j+1})\,\varrho\langle\prod_jv_j^{m_j}\rangle.
\end{equation}
%
In other word, all the velocity moments are
essentially the coefficients of the MacLaurin--Taylor series expansion
of $\varphi$ (at $\bm k=0$) and vice versa.
Moreover, if all velocity moments with an odd power of $v_1$ vanish,
then $\varphi$ is symmetric under the transform $k^1\leftrightarrow-k^1$,
for all the coefficients in the MacLaurin series for the odd-power
terms of $k^1$ vanish. On the other hand, the same symmetry for
the real part $\Re\varphi$ (which is also the even part if $\df$ is real)
of $\varphi$ is similarly deduced only with even velocity moments.
Finally the distribution function is recovered through
the inverse Fourier transform,
\begin{equation}\begin{split}
(2\pi)^n\df&={\textstyle\idotsint}dk^1\dotsm dk^n
e^{-{\rm i}\bm{p\cdot k}}\varphi(\bm k);
\\\textstyle
(2\pi)^n\df^+&={\textstyle\idotsint}dk^1\dotsm dk^n
\cos(\bm{p\cdot k})\,\Re\varphi(\bm k),
\end{split}\end{equation}
where $\df$ is assumed to be real. Then
$\df(-p_1,\dotsc,p_n)=\int\!d^n\!\bm k\,
e^{{\rm i}(p_1k^1-\sum_{j=2}^np_jk^j)}\varphi(k^1,\dotsc,k^n)
=\int\!d^n\!\bm k\,e^{-{\rm i}\bm{p\cdot k}}\varphi(-k^1,\dotsc,k^n)$
and $\df^+(-p_1,\dotsc,p_n)
=\int\!d^n\!\bm k\,\cos(\bm{p\cdot k})\Re\varphi(-k^1,\dotsc,k^n)$.
Thus, if $\varphi$ is symmetric under $k^1\leftrightarrow-k^1$,
then $\df$ is also symmetric under $p_1\leftrightarrow-p_1$,
whereas the even part $\df^+$ is symmetric under $p_1\leftrightarrow-p_1$
if the real part $\Re\varphi$ of
the characteristic function is symmetric under $k^1\leftrightarrow-k^1$,
which completes the proof of Lemma \ref{lem}.
Since the argument is valid irrespective of the label for the index,
Corollary \ref{corol-df} and Lemma \ref{lem} together then imply
Corollary \ref{corol-vm}.

\section{The Jeans equations}
\label{sec:je}

Corollary \ref{corol-vm} is still of little practical use as it refers to
the infinite set of all of even velocity moments, which is difficult
to constrain from observables. Instead, we would like to seek the
sufficient condition for the St\"ackel potential referring only to a
finite subset of the velocity moments. For this, we need to establish
the explicit relations among the velocity moments of the system in
equilibrium first. This may be achieved through taking the moment
integrals of the CBE \citep{BT}. The resulting first moment equations
correspond to the usual Jeans equations. Here, we
derive all of the $m$-th moment equations in an arbitrary coordinate system.

\subsection{In an arbitrary coordinate system}

Suppose that $(q^1,\dotsc,q^n)$ is an arbitrary coordinate with
$g_{\mu\nu}$ being its metric coefficient (so that the line element is
$ds^2=g_{\mu\nu}dq^\mu dq^\nu$; throughout this section, the Einstein
summation convention for \emph{Greek} indices are assumed).
Let us think of the Hamiltonian of the form
$\ham=\frac12g^{\mu\nu}p_\mu p_\nu+\Phi(q^1,\dotsc,q^n)$,
where $g^{\mu\nu}$ is the inverse metric.
Then the CBE (assuming $\partial\df/\partial t=0$)
in the canonical phase-space coordinate is equivalent to
\begin{equation}\label{eq:cbe}
\left\lbrace\df,\ham\right\rbrace
=g^{\mu\nu}p_\nu\frac{\partial\df}{\partial q^\mu}
-\left(\frac12\frac{\partial g^{\lambda\zeta}}{\partial q^\mu}p_\lambda p_\zeta
+\frac{\partial\Phi}{\partial q^\mu}\right)
\frac{\partial\df}{\partial p_\mu}=0.
\end{equation}
Next consider integrating this over the momentum space after
multiplying by $\prod_ip_{\iota_i}$, where $(\iota_1,\dotsc)$
is a sequence of indices with $\iota_i\in\set{1,\dotsc,n}$,
and also utilizing integration by parts
\begin{multline}
\int\!d^n\!\bm p\,\frac{\partial\df}{\partial p_\nu}\prod_jp_{\iota_j}
=\cancel{\int\!d^n\!\bm p\,
\frac\partial{\partial p_\nu}\left(\df\prod_jp_{\iota_j}\right)}
\\-\int\!d^n\!\bm p\,\df
\frac{\partial(\prod_jp_{\iota_j})}{\partial p_\nu},
\end{multline}
where $d^n\!\bm p\equiv dp_1\dotsm dp_n$. Since
$(q^1,\dotsc,q^n,p_1,\dotsc,p_n)$ is a canonical phase-space coordinate,
$dq^1\dotsm dq^ndp_1\dotsm dp_n=d^n\!\bm x\,d^n\!\bm v$
where $d^n\!\bm x$ and $d^n\!\bm v$ are the volume $n$-forms
for the configuration and the velocity spaces.
However $d^n\!\bm x=h\,dq^1\dotsm dq^n$ where $h^2\equiv\det(g_{ij})$,
and so this gives the Jacobian determinant as $d^n\!\bm p=h\,d^n\!\bm v$.
Therefore
\begin{equation}\textstyle
\int\!d^n\!\bm p\,\df\prod_ip_{\iota_i}
=h\int\!d^n\!\bm v\,\df\prod_ip_{\iota_i}
=h\,\varrho\,\langle\prod_ip_{\iota_i}\rangle,
\end{equation}
and the $m$-th moment integrals of equation (\ref{eq:cbe})
(divided by $h\varrho$) result in
\begin{multline}\label{eq:cbem}
\sum_{i=1}^m\frac{\partial\Phi}{\partial q^{\iota_i}}
V^{(m-1)}_{\iota_1\dotsm\cancel{\iota_i}\dotsm\iota_m}
+\frac{g^{\mu\nu}}{h\varrho}\frac{\partial}{\partial q^\mu}
\left(h\varrho V^{(m+1)}_{\nu\iota_1\dotsm\iota_m}\right)
\\+\frac{\partial g^{\lambda\zeta}}{\partial q^\lambda}
V^{(m+1)}_{\zeta\iota_1\dotsm\iota_m}
+\frac12\sum_{i=1}^m\frac{\partial g^{\lambda\zeta}}{\partial q^{\iota_i}}
V^{(m+1)}_{\lambda\zeta\iota_1\dotsm\cancel{\iota_i}\dotsm\iota_m}=0
\end{multline}
where the slash through the index
represents skipping the particular index, while
$V^{(m)}_{\iota_1\dotsm\iota_m}=\langle p_{\iota_1}\dotsm
p_{\iota_m}\rangle$ is the $m$-th momentum moment, which forms a
symmetric $(0,m)$-tensor.

Thanks to the relation \citep[see][\S~2.11]{AW05}
\begin{equation}
\frac{\partial\ln h^2}{\partial q^\mu}
=g^{\lambda\zeta}\frac{\partial g_{\lambda\zeta}}{\partial q^\mu}
\end{equation}
equation (\ref{eq:cbem}) is further reducible to
\begin{multline}\label{eq:cbep}
\sum_{i=1}^m\frac{\partial\Phi}{\partial q^{\iota_i}}
V^{(m-1)}_{\iota_1\dotsm\cancel{\iota_i}\dotsm\iota_m}
+\frac1\varrho\frac\partial{\partial q^\mu}
\left(\varrho\tilde V^\mu_{\iota_1\dotsm\iota_m}\right)
\\+\frac{g^{\lambda\zeta}}2\frac{\partial g_{\lambda\zeta}}{\partial q^\mu}
\tilde V^\mu_{\iota_1\dotsm\iota_m}
-\sum_{i=1}^m\frac{g^{\lambda\zeta}}2
\frac{\partial g_{\lambda\mu}}{\partial q^{\iota_i}}
\tilde V^\mu_{\zeta\iota_1\dotsm\cancel{\iota_i}\dotsm\iota_m}=0
\end{multline}
where $\tilde V^\mu_{\iota_1\dotsm\iota_m}
=\langle\dot q^\mu p_{\iota_1}\dotsm p_{\iota_m}\rangle$,
which is basically the same tensor as $V^{(m+1)}$ but one of the index raised,
$\tilde V^\mu_{\iota_1\dotsm\iota_m}
=g^{\mu\nu}V^{(m+1)}_{\nu\iota_1\dotsm\iota_m}$. Also used are
$g_{\lambda\mu}(\partial g^{\lambda\zeta}/\partial q^\iota)
=-g^{\lambda\zeta}(\partial g_{\lambda\mu}/\partial q^\iota)$, which
follows $g^{\lambda\zeta}g_{\lambda\mu}=\delta^\lambda_\mu$.
Utilizing the covariant derivative of the tensor
\begin{equation}
\nabla_\mu T^{\lambda_1\dotsm}_{\zeta_1\dotsm}
=\frac{\partial T^{\lambda_1\dotsm}_{\zeta_1\dotsm}}{\partial q^\mu}
+\sum_j\Gamma^{\lambda_j}_{\mu\nu}
T^{\lambda_1\dotsm\cancel{\lambda_j}\nu\dotsm}_{\zeta_1\dotsm}
-\sum_i\Gamma^\nu_{\mu\zeta_i}
T^{\lambda_1\dotsm}_{\zeta_1\dotsm\cancel{\zeta_i}\nu\dotsm}
\end{equation}
defined with the Levi-Civita connection coefficients
\begin{equation}\label{eq:mcc}
\Gamma^\mu_{\lambda\zeta}
=\frac{g^{\mu\nu}}2\left(
\frac{\partial g_{\lambda\nu}}{\partial q^\zeta}
+\frac{\partial g_{\nu\zeta}}{\partial q^\lambda}
-\frac{\partial g_{\lambda\zeta}}{\partial q^\nu}\right),
\end{equation}
equation (\ref{eq:cbep}) finally simplifies to 
\begin{equation}\label{eq:jeans}
\nabla_\mu
\left(\varrho\langle\dot q^\mu p_{\iota_1}\dotsm p_{\iota_m}\rangle\right)
+\varrho\sum_{(\iota_i)}\frac{\partial\Phi}{\partial q^{\iota_1}}
\langle p_{\iota_2}\dotsm p_{\iota_m}\rangle=0,
\end{equation}
where the sum is over all cyclic permutations through the indices. Here, the
$\nabla_\mu$-term is in fact the divergence of the $(1,m)$-tensor
field $\langle\dot q^\mu p_{\iota_1}\dotsm p_{\iota_m}\rangle$, which,
given $\nabla_\mu g^{\lambda\zeta}=0$, is also equivalent to
$\nabla_\mu(\varrho\langle\dot q^\mu p_{\iota_1}\dotsm p_{\iota_m}\rangle)
=g^{\mu\nu}\nabla_\mu
(\varrho\langle p_\nu p_{\iota_1}\dotsm p_{\iota_m}\rangle)$.

Since equation (\ref{eq:jeans}) is symmetric with respect to any
permutation of free indices among $\set{\iota_1,\dotsc,\iota_m}$,
there are $\left(\!\binom nm\!\right)=\binom{n+m-1}m=(n)_m/m!$
independent equations for a fixed $m$ in $n$ dimensions
-- here $\left(\!\binom nm\!\right)$ is the $m$-combination out of
$n$ elements \emph{with} repetition, and $(n)_m\equiv\prod_{j=0}^{m-1}(n+j)$
is the rising sequential product. The single equation for
$m=0$; that is, $\nabla_\mu(\varrho\langle\dot q^\mu\rangle)=0$,
is simply the continuity equation $\bm\nabla\bm\cdot(\varrho\bm v)=0$
for the time-independent density field, whereas the $m=1$ equation,
$\nabla_\mu(\varrho\langle\dot q^\mu p_\nu\rangle)
+\varrho(\partial\Phi/\partial q^\nu)=0$, basically corresponds to
the static Euler equation with an anisotropic stress
tensor $\mathbf P$ (i.e.\ the Cauchy or Navier--Stokes
momentum equation in fluid mechanics or the Jeans equation in stellar
dynamics); namely $\bm\nabla\bm\cdot\mathbf P+\varrho\bm\nabla\Phi=0$.

\subsection{In an orthogonal coordinate}

In an orthogonal coordinate with scale factors $h_i$, the metric is
diagonal as in $g_{ij}=0$ for $i\ne j$ and $g_{ii}=h_i^2$. Since the
velocity component $v_i$ projected onto the \emph{orthonormal} frame
is related to the specific momentum component via $p_j=h_jv_j$ (and
$v_j=h_j\dot q^j$), the tensor components in the orthogonal
coordinates are related to the orthogonal velocity moments through
\begin{equation}
\langle v_jv_{\ell_1}\dotsm v_{\ell_m}\rangle
=\frac{V^{(m+1)}_{j\ell_1\dotsm\ell_m}}{h_jh_{\ell_1}\dotsm h_{\ell_m}}
=\frac{h_j\tilde V^j_{\ell_1\dotsm\ell_m}}{h_{\ell_1}\dotsm h_{\iota_m}}.
\end{equation}
Then equation (\ref{eq:cbem}) or (\ref{eq:cbep}) reduces to
(here $h=\prod_jh_j$)
\begin{equation}\begin{split}
\sum_{i=1}^m\varrho&
\left\langle\frac{v_{\ell_1}\dotsm v_{\ell_m}}{v_{\ell_i}}\right\rangle
\frac{\partial\Phi}{h_{\ell_i}\partial q^{\ell_i}}
+\sum_{j=1}^n\frac{\partial}{h_j\partial q^j}\left(
\varrho\langle v_j v_{\ell_1}\dotsm v_{\ell_m}\rangle\right)
\\&+\sum_{j=1}^n\varrho\langle v_jv_{\ell_1}\dotsm v_{\ell_m}\rangle
\frac\partial{h_j\partial q^j}\left[\ln
\left(\frac{hh_{\ell_1}\dotsm h_{\ell_m}}{h_j}\right)\right]
\\&-\sum_{i=1}^m\sum_{j=1}^n\varrho
\left\langle v_j^2\frac{v_{\ell_1}\dotsm v_{\ell_m}}{v_{\ell_i}}\right\rangle
\frac{\partial\ln h_j}{h_{\ell_i}\partial q^{\ell_i}}=0
\end{split}\end{equation}
in the orthogonal coordinate. For $m=0$, this becomes
\begin{equation}
\sum_{j=1}^n
\left[\frac{\partial\varrho\langle v_j\rangle}{h_j\partial q^j}
+\varrho\langle v_j\rangle\frac{\partial\ln(h/h_j)}{h_j\partial q^j}\right]=0.
\end{equation}
The $m=1$ case results in the Jeans equation
\begin{equation}\label{eq:jeans2}
\sum_{j=1}^n\left[\frac1\varrho
\frac{\partial\varrho\langle v_j v_\ell\rangle}{h_j\partial q^j}
+\langle v_jv_\ell\rangle
\frac{\partial\ln(hh_\ell/h_j)}{h_j\partial q^j}
-\langle v_j^2\rangle\frac{\partial\ln h_j}{h_\ell\partial q^\ell}\right]
=-\frac{\partial\Phi}{h_{\ell}\partial q^{\ell}}
\end{equation}
with a fixed $\ell\in\set{1,\dotsc,n}$.
The Jeans equations in an arbitrary 3-dimensional curvilinear coordinate
system have been derived before \citep{LB60,ELB},
although the usual expressions typically involve the second moments
decomposed into those due to random and coherent motions; namely,
$\langle v_j v_\ell\rangle=
\langle v_j\rangle\langle v_\ell\rangle+\sigma_{j\ell}^2$ etc.

For our purpose here, we also require the expression for the $m=3$
equations: with fixed $j,k,\ell$
\begin{multline}\label{eq:jeans4}
\sum_{i=1}^n\Biggl[\frac1\varrho
\frac{\partial\varrho\langle v_iv_jv_kv_\ell\rangle}{h_i\partial q^i}
+\langle v_iv_jv_kv_\ell\rangle
\frac{\partial\ln(hh_jh_kh_\ell/h_i)}{h_i\partial q^i}
\\-\langle v_i^2v_kv_\ell\rangle
\frac{\partial\ln h_i}{h_j\partial q^j}
-\langle v_i^2v_jv_\ell\rangle
\frac{\partial\ln h_i}{h_k\partial q^k}
-\langle v_i^2v_jv_k\rangle
\frac{\partial\ln h_i}{h_\ell\partial q^\ell}\Biggr]
\\=-\left(\langle v_kv_\ell\rangle\frac{\partial\Phi}{h_j\partial q^j}
+\langle v_jv_\ell\rangle\frac{\partial\Phi}{h_k\partial q^k}
+\langle v_jv_k\rangle\frac{\partial\Phi}{h_\ell\partial q^\ell}\right).
\end{multline}

\section{The second and fourth moments\\in the St\"ackel potentials}
\label{sec:sfmsp}

Now we are ready to prove the main finding:
\begin{theorem}\label{theo-ms}
Suppose in the St\"ackel coordinate $(q^1,\dotsc,q^n)$ that
all mixed second moments of the steady-state tracer velocities
vanish and the remaining second moments are all distinct
(i.e.\ $\langle v_iv_j\rangle=0$ and
$\langle v_i^2\rangle\ne\langle v_j^2\rangle$ for all $i\ne j$)
and only non-vanishing fourth velocity moments of the tracers are
those in the form of $\langle v_i^4\rangle$ or $\langle v_i^2v_j^2\rangle$.
Then the potential must be separable in the given St\"ackel coordinate.
\end{theorem}
{\it Proof:}
Under the given condition, equation (\ref{eq:jeans2}) simplifies to
\begin{equation}\label{eq:j2}
\frac{\partial\varrho\langle v_j^2\rangle}{\partial q^j}
+\varrho\langle v_j^2\rangle\frac{\partial\ln h}{\partial q^j}
-\sum_{i=1}^n\varrho\langle v_i^2\rangle\frac{\partial\ln h_i}{\partial q^j}
+\varrho\frac{\partial\Phi}{\partial q^j}=0,
\end{equation}
while equations (\ref{eq:jeans4}) with $j=k=\ell$ reduce to
\begin{equation}\label{eq:j4s}
\frac{\partial\varrho\langle v_j^4\rangle}{\partial q^j}
+\varrho\langle v_j^4\rangle
\frac{\partial\ln(hh_j^2)}{\partial q^j}
-\sum_{i=1}^n\varrho
\langle v_i^2v_j^2\rangle\frac{\partial\ln h_i^3}{\partial q^j}
+3\varrho\langle v_j^2\rangle\frac{\partial\Phi}{\partial q^j}=0;
\end{equation}
and those with $j\ne k=\ell$ to
\begin{equation}\label{eq:j4d}
\frac{\partial\varrho\langle v_j^2v_k^2\rangle}{\partial q^j}
+\varrho\langle v_j^2v_k^2\rangle
\frac{\partial\ln(hh_k^2)}{\partial q^j}
-\sum_{i=1}^n\varrho
\langle v_i^2v_k^2\rangle\frac{\partial\ln h_i}{\partial q^j}
+\varrho\langle v_k^2\rangle\frac{\partial\Phi}{\partial q^j}=0.
\end{equation}
Differentiating equation (\ref{eq:j4d}) with respect to $q^k$ results in
\begin{multline}\label{eq:j4der}
\frac{\partial^2\varrho\langle v_j^2v_k^2\rangle}{\partial q^k\partial q^j}
+\frac{\partial\varrho\langle v_j^2v_k^2\rangle}{\partial q^k}
\frac{\partial\ln(hh_k^2)}{\partial q^j}
+\varrho\langle v_j^2v_k^2\rangle
\frac{\partial^2\ln(hh_k^2)}{\partial q^k\partial q^j}
\\-\sum_{i=1}^n\left[
\frac{\partial\varrho\langle v_i^2v_k^2\rangle}{\partial q^k}
\frac{\partial\ln h_i}{\partial q^j}+\varrho\langle v_i^2v_k^2\rangle
\frac{\partial^2\ln h_i}{\partial q^k\partial q^j}\right]
\\+\frac{\partial\varrho\langle v_k^2\rangle}{\partial q^k}
\frac{\partial\Phi}{\partial q^j}
+\varrho\langle v_k^2\rangle\frac{\partial^2\Phi}{\partial q^k\partial q^j}=0
\quad(j\ne k).
\end{multline}
Note that the same equation with indices $j\leftrightarrow k$ switched
also holds. Hence the second derivative term
$\partial^2(\varrho\langle v_j^2v_k^2\rangle)/(\partial q^k\partial q^j)$,
which is symmetric under $j\leftrightarrow k$, can be eliminated by
subtracting this from the $j\leftrightarrow k$ switched equation: that is,
\begin{multline}
\frac{\partial\varrho\langle v_j^2v_k^2\rangle}{\partial q^k}
\frac{\partial\ln(hh_k^2/h_j)}{\partial q^j}
-\frac{\partial\varrho\langle v_j^2v_k^2\rangle}{\partial q^j}
\frac{\partial\ln(hh_j^2/h_k)}{\partial q^k}
\\+\frac{\partial\varrho\langle v_j^4\rangle}{\partial q^j}
\frac{\partial\ln h_j}{\partial q^k}
-\frac{\partial\varrho\langle v_k^4\rangle}{\partial q^k}
\frac{\partial\ln h_k}{\partial q^j}
+\varrho\langle v_j^2v_k^2\rangle
\frac{\partial^2\ln(h_k^2/h_j^2)}{\partial q^j\partial q^j}
\\+\sum_{\substack{i=1\\i\ne j,k}}^n
\left[\frac{\partial\varrho\langle v_i^2v_j^2\rangle}{\partial q^j}
\frac{\partial\ln h_i}{\partial q^k}
-\frac{\partial\varrho\langle v_i^2v_k^2\rangle}{\partial q^k}
\frac{\partial\ln h_i}{\partial q^j}\right]
\\+\sum_{i=1}^n\varrho
\left(\langle v_i^2v_j^2\rangle-\langle v_i^2v_k^2\rangle\right)
\frac{\partial^2\ln h_i}{\partial q^j\partial q^k}
\\=\frac{\partial\varrho\langle v_j^2\rangle}{\partial q^j}
\frac{\partial\Phi}{\partial q^k}
-\frac{\partial\varrho\langle v_k^2\rangle}{\partial q^k}
\frac{\partial\Phi}{\partial q^j}
+\varrho(\langle v_j^2\rangle-\langle v_k^2\rangle)
\frac{\partial^2\Phi}{\partial q^j\partial q^k},
\end{multline}
where the remaining spatial derivatives of the moments
can be replaced by means of equations (\ref{eq:j2}),
(\ref{eq:j4s}) and (\ref{eq:j4d}). After tedious but trivial algebra,
we then obtain
\begin{multline}\label{eq:main}
(\langle v_j^4\rangle-3\langle v_j^2v_k^2\rangle)
h_j^2\dst_{jk}(h_j^{-2})
-(\langle v_k^4\rangle-3\langle v_j^2v_k^2\rangle)
h_k^2\dst_{jk}(h_k^{-2})
\\+\sum_{\substack{i=1\\i\ne j,k}}^n
(\langle v_i^2v_j^2\rangle-\langle v_i^2v_k^2\rangle)
h_i^2\dst_{jk}(h_i^{-2})
\\+2(\langle v_j^2\rangle-\langle v_k^2\rangle)
\dst_{jk}(\Phi)=0,
\end{multline}
where $\dst_{jk}$ is as defined in equation (\ref{eq:ddef}).
In the St\"ackel coordinate such that $\dst_{jk}(h_i^{-2})=0$ for any
$j\ne k$ and all $i$, equation (\ref{eq:main}) then indicates
$(\langle v_j^2\rangle-\langle v_k^2\rangle)\dst_{jk}(\Phi)=0$. So
given $\langle v_j^2\rangle\ne\langle v_k^2\rangle$ for all $j\ne k$,
the potential must be separable in the given St\"ackel
coordinate.

\section{Partially separable potentials\\in 3-dimensional space}
\label{sec:psp3d}

Theorem \ref{theo-ms} provides a sufficient condition for the potential
to be separable in the given St\"ackel coordinate in terms of the
second and fourth velocity moments of the tracers. In 3-dimensional space,
the separable potential satisfies
$\dst_{12}(\Phi)=\dst_{13}(\Phi)=\dst_{23}(\Phi)=0$. However
$\dst_{23}(\Phi)=0$ does not explicitly involve $q^1$ and so one might expect
that the condition $\dst_{12}(\Phi)=\dst_{13}(\Phi)=0$ may be implied
by only those moments involving $v_1$. In fact, we can establish: 
\begin{theorem}\label{theo-ps}
Let $(q^1,q^2,q^3)$ be the St\"ackel coordinate with the scale
factors $\set{h_1,h_2,h_3}$ satisfying $(\partial/\partial q^1)(h_2/h_3)=0$.
If all the second and fourth velocity moments
of the steady-state tracers with an odd power to $v_1$ vanish
(i.e.\ $\langle v_1v_2\rangle=\langle v_1v_3\rangle=0$,
$\langle v_1^3v_2\rangle=\langle v_1^3v_3\rangle=0$ and
$\langle v_1v_2^3\rangle=\langle v_1v_2^2v_3\rangle
=\langle v_1v_2v_3^2\rangle=\langle v_1v_3^2\rangle=0$),
and $(\langle v_1\rangle-\langle v_2\rangle)
(\langle v_1\rangle-\langle v_3\rangle)\ne\langle v_2v_3\rangle^2$,
then the potential satisfies the partial differential
equations $\dst_{12}(\Phi)=\dst_{13}(\Phi)=0$,
where $\dst_{ij}(\Phi)$ is as defined in equation (\ref{eq:ddef}).
\end{theorem}
Here we provide only a sketch of the proof.
First, consider equation (\ref{eq:jeans4}) with
$\{j,k,\ell\}=\{1,2,3\}$, $\{1,2,2\}$, $\{1,1,2\}$
under the given conditions:
\begin{gather}\label{eq:j123}\begin{split}
\frac{\partial\varrho\langle v_1^2v_2v_3\rangle}{\partial q^1}
&+\varrho\langle v_1^2v_2v_3\rangle
\frac{\partial\ln(h_2^2h_3^2)}{\partial q^1}
-\varrho\langle v_2^3v_3\rangle\frac{\partial\ln h_2}{\partial q^1}
\\&-\varrho\langle v_2v_3^3\rangle\frac{\partial\ln h_3}{\partial q^1}
+\varrho\langle v_2v_3\rangle\frac{\partial\Phi}{\partial q^1}=0;
\end{split}\\\label{eq:j122}\begin{split}
\frac{\partial\varrho\langle v_1^2v_2^2\rangle}{\partial q^1}
&+\varrho\langle v_1^2v_2^2\rangle\frac{\partial\ln(h_2^3h_3)}{\partial q^1}
-\varrho\langle v_2^4\rangle\frac{\partial\ln h_2}{\partial q^1}
\\&-\varrho\langle v_2^2v_3^2\rangle\frac{\partial\ln h_3}{\partial q^1}
+\varrho\langle v_2^2\rangle\frac{\partial\Phi}{\partial q^1}=0;
\end{split}\\\label{eq:j112}\begin{split}
\frac{\partial\varrho\langle v_1^2v_2^2\rangle}{\partial q^2}
&+\varrho\langle v_1^2v_2^2\rangle\frac{\partial\ln(h_1^3h_3)}{\partial q^2}
-\varrho\langle v_1^2v_3^2\rangle\frac{\partial\ln h_3}{\partial q^2}
\\&+\frac{h_2}{h_3}\left[
\frac{\partial\varrho\langle v_1^2v_2v_3\rangle}{\partial q^3}
+\varrho\langle v_1^2v_2v_3\rangle
\frac{\partial\ln(h_1^3h_2^2)}{\partial q^3}\right]
\\&-\varrho\langle v_1^4\rangle\frac{\partial\ln h_1}{\partial q^2}
+\varrho\langle v_1^2\rangle\frac{\partial\Phi}{\partial q^2}=0.
\end{split}\end{gather}
Among the partial derivatives of
equation (\ref{eq:j123}) with respect to $q^3$,
equation (\ref{eq:j122}) with respect to $q^2$ and
equation (\ref{eq:j112}) with respect to $q^1$,
the two second derivatives of the fourth moments,
$\partial^2(\varrho\langle v_1^2v_2^2\rangle)/(\partial q^1\partial q^2)$
and $\partial^2(\varrho\langle v_1^2v_2v_3\rangle)/(\partial q^1\partial q^3)$
can be eliminated, which leaves a single equation
relating the second and fourth moments and their first derivatives.
It turns out all the first derivatives
in the resulting equation can be replaced
by means of the Jeans equations (\ref{eq:jeans2}) and (\ref{eq:jeans4}),
except $\partial(\varrho\langle v_1^2v_2v_3\rangle)/(\partial q^3)$.
Following lengthy algebra, we arrive at
\begin{multline}\label{eq:jm}
A_1h_1^2\dst_{12}(h_1^{-2})-A_2h_2^2\dst_{12}(h_2^{-2})+Bh_3^2\dst_{12}(h_3^{-2})
\\-\left[3\langle v_1^2v_2v_3\rangle h_1^2\dst_{13}(h_1^{-2})
+\langle v_2^3v_3\rangle h_2^2\dst_{13}(h_2^{-2})
+Ch_3^2\dst_{13}(h_3^{-2})\right]\hbar
\\+\frac4\varrho\left[
\frac{\partial\varrho\langle v_1^2v_2v_3\rangle}{\partial q^3}
+\varrho\langle v_1^2v_2v_3\rangle\frac{\partial\ln h_1^3h_2^2}{\partial q^3}
\right]\frac{\partial\hbar}{\partial q^1}
\\
+2\left[(\langle v_1^2\rangle-\langle v_2^2\rangle)\dst_{12}(\Phi)
-\langle v_2v_3\rangle\hbar\,\dst_{13}(\Phi)\right]=0.
\end{multline}
where
$A_1\equiv\langle v_1^4\rangle-3\langle v_1^2v_2^2\rangle$,
$A_2\equiv\langle v_2^4\rangle-3\langle v_1^2v_2^2\rangle$,
$B\equiv\langle v_1^2v_3^2\rangle-\langle v_2^2v_3^2\rangle$, and
$C\equiv\langle v_2v_3^3\rangle-2\langle v_1^2v_2v_3\rangle$,
while $\hbar\equiv h_2/h_3$. It is obvious that the same equation with
the indices $2\leftrightarrow3$ switched holds too.
In the St\"ackel coordinate with $(\partial/\partial q^1)(h_2/h_3)=0$,
these two then simplify to
\begin{equation}\label{eq:theo-ps}\begin{split}
h_3(\langle v_1^2\rangle-\langle v_2^2\rangle)\dst_{12}(\Phi)
&=h_2\langle v_2v_3\rangle\dst_{13}(\Phi);
\\h_2(\langle v_1^2\rangle-\langle v_3^2\rangle)\dst_{13}(\Phi)
&=h_3\langle v_2v_3\rangle\dst_{12}(\Phi).
\end{split}\end{equation}
Provided that $(\langle v_1\rangle-\langle v_2\rangle)
(\langle v_1\rangle-\langle v_3\rangle)\ne\langle v_2v_3\rangle^2$,
they are linearly independent and so imply
$\dst_{12}(\Phi)=\dst_{13}(\Phi)=0$.

The importance of this result becomes clearer with a concrete choice
of the St\"ackel coordinate. In particular,
\begin{corol}
If the steady-state tracer velocity moments
in the Cartesian coordinate $(x,y,z)$
are constrained such that $\langle v_xv_z\rangle=\langle v_yv_z\rangle=0$,
$\langle v_xv_z^3\rangle=\langle v_yv_z^3\rangle=0$
and $\langle v_x^3v_z\rangle=\langle v_x^2v_yv_z\rangle=
\langle v_xv_y^2v_z\rangle=\langle v_y^3v_z\rangle=0$,
the potential must be in the form of $\Phi(x,y,z)=f(x,y)+g(z)$,
provided that $(\langle v_x^2\rangle-\langle v_z^2\rangle)
(\langle v_y^2\rangle-\langle v_z^2\rangle)\ne\langle v_xv_y\rangle^2$.
\end{corol}
The scale factors of the Cartesian coordinate
are $h_x=h_y=h_z=1$ and so $\partial(h_x/h_y)/\partial z=0$.
By Theorem \ref{theo-ps}, the condition then implies
$\dst_{xz}(\Phi)=\partial^2\Phi/(\partial x\partial z)=0$
and $\dst_{yz}(\Phi)=\partial^2\Phi/(\partial y\partial z)=0$;
that is, $\partial\Phi/\partial z$ is a function of $z$ alone
and so its general solution is $\Phi(x,y,z)=f(x,y)+g(z)$.

It is clear that a similar result also holds for any 3-dimensional
St\"ackel coordinate that is translation-symmetric along the
$z$-direction (including the cylindrical-polar, elliptic-cylindrical,
and parabolic-cylindrical coordinates). In fact, this is true even
for any 3-dimensional coordinate resulting from the linear
duplication of a 2-dimensional coordinate (not necessarily
St\"ackel). That is to say, if $(q^1,q^2,z)$ is an orthogonal
coordinate for the 3-dimensional Euclidean space such that the
coordinate surfaces for a fixed $z$ are parallel planes with
$(q^1,q^2)$ being a translationally invariant coordinate system on
each of the planes, then the scale factors $h_1$ and $h_2$ must be
independent of $z$, while $h_z$ is a function of $z$ alone (which can
always be set to the unity after rescaling). Therefore
$\dst_{iz}=\partial^2/(\partial q^i\partial z)$ for $i\in\set{1,2}$ in such a
coordinate and so it follows that
$\dst_{iz}(h_j^{-2})=\dst_{iz}(h_z^{-2})=0$. Since
$\partial(h_1/h_2)/\partial z=0$, if all second and fourth moments
with the odd power to $v_z$ vanish, equation (\ref{eq:jm}) in this
coordinate still implies equation (\ref{eq:theo-ps}), and thus the
potential must satisfy $\partial^2\Phi/(\partial q^1\partial z)
=\partial^2\Phi/(\partial q^2\partial z)=0$; that is, the potential
being decomposable into a function of the height alone and that of the
mid-plane coordinates; namely $\Phi=f(q^1,q^2)+g(z)$.

Separable potentials in Cartesians are rather unrealistic, and so we
turn to the condition for axisymmetric potentials,
\begin{corol}
The steady-state tracer population with
$\langle v_Rv_\phi\rangle=\langle v_zv_\phi\rangle=0$,
$\langle v_Rv_\phi^3\rangle=\langle v_zv_\phi^3\rangle=0$
and $\langle v_R^3v_\phi\rangle=\langle v_R^2v_zv_\phi\rangle=
\langle v_Rv_z^2v_\phi\rangle=\langle v_z^3v_\phi\rangle=0$
in the cylindrical-polar coordinate $(R,\phi,z)$
implies that the potential must be in the form of
$\Phi(R,\phi,z)=R^{-2}f(\phi)+g(R,z)$,
provided that $(\langle v_R^2\rangle-\langle v_\phi^2\rangle)
(\langle v_z^2\rangle-\langle v_\phi^2\rangle)\ne\langle v_Rv_z\rangle^2$.
\end{corol}
The scale factors of the cylindrical-polar coordinate
is $h_R=h_z=1$ and $h_\phi=R$, and so $\partial(h_R/h_z)/\partial\phi=0$.
According to Theorem \ref{theo-ps}, the condition implies that
\begin{equation}\begin{split}
\dst_{R\phi}(\Phi)
&=\frac{\partial^2\Phi}{\partial R\partial\phi}
+\frac2R\frac{\partial\Phi}{\partial\phi}
=\frac1{R^2}
\frac\partial{\partial R}\left(R^2\frac{\partial\Phi}{\partial\phi}\right)=0;
\\\dst_{\phi z}(\Phi)
&=\frac{\partial^2\Phi}{\partial\phi\partial z}
=\frac1{R^2}
\frac\partial{\partial z}\left(R^2\frac{\partial\Phi}{\partial\phi}\right)=0.
\end{split}\end{equation}
That is to say, $R^2(\partial\Phi/\partial\phi)=F(\phi)$ is a function of
$\phi$ alone. Integrating $R^{-2}F(\phi)$ over $\phi$, the general
solution is therefore of the form $\Phi(R,\phi,z)=R^{-2}f(\phi)+g(R,z)$.

Provided that $\Phi(R,\phi,z)$ is single-valued, $f(\phi)$ must be
$2\pi$-periodic. If $f(\phi)=\sum_ka_k\cos[k(\phi-\phi_k)]$ is the Fourier
series expansion, the density profile for $\Phi\propto f/R^2$ behaves
like
\begin{equation}
\nabla^2\left(\frac f{R^2}\right)
=\frac1{R^4}\sum_k(4-k^2)\,a_k\cos[k(\phi-\phi_k)].
\end{equation}
Unless $a_k=0$ for all $k\ne2$, this is unintegrable as $R\to0$,
which is considered unphysical.
The case $f(\phi)\propto\cos[2(\phi-\phi_0)]$ is technically allowed
but this potential is only due to the choice of the boundary condition
and no actual source in otherwise empty space can generate such a potential.
Hence we may in fact further infer that
the potential resulting from the corollary is axisymmetric (i.e.\ $f=0$).

Similar to the translation symmetric cases, the result is in fact
applicable for any 3-dimensional coordinate constructed by
rotating a reflection-symmetric 2-dimensional coordinate along
its symmetry axis
(e.g., the cylindrical-polar, spherical-polar, rotational-parabolic,
and oblate and prolate spheroidal coordinates). In particular,
in the 3-dimensional coordinate $(q^1,q^2,\phi)$ with the scale factors
$h_1$ and $h_2$ that are independent of $\phi$ -- so
$\partial(h_1/h_2)/\partial\phi=0$ -- and $h_\phi=R(q^1,q^2)$, we have
\begin{equation}
\dst_{i\phi}(f)=\frac{\partial^2f}{\partial q^i\partial\phi}
+\frac{\partial\ln R^2}{\partial q^i}\frac{\partial f}{\partial\phi}
=\frac1{R^2}\frac\partial{\partial q^i}
\left(R^2\frac{\partial f}{\partial\phi}\right)
\end{equation}
for $i\in\set{1,2}$.
Given that all scale factors are assumed to be independent of $\phi$,
if all second and fourth moments with the odd power to $v_\phi$ vanish,
we then infer that $F\equiv R^2(\partial\Phi/\partial\phi)$ is
a function of $\phi$ alone
(i.e.\ $\partial F/\partial q^1=\partial F/\partial q^2=0$)
and subsequently the potential must be axisymmetric.

Here we also note that the vanishing moments are the true moments but
not the central moments (such as the co-variance or co-kurtosis etc.),
which may be seen by the fact that
the underlying potential only determines the orbit of individual tracers
and does not distinguish between coherent and random motions for groups of
tracers.

Lastly, we consider the case of spherical-polar coordinates for which
we have
\begin{corol}
The steady-state tracer population with
$\langle v_rv_\theta\rangle=\langle v_rv_\phi\rangle=0$,
$\langle v_r^3v_\theta\rangle=\langle v_r^3v_\phi\rangle=0$
and $\langle v_rv_\theta^3\rangle=\langle v_rv_\theta^2v_\phi\rangle=
\langle v_rv_\theta v_\phi^2\rangle=\langle v_rv_\phi^3\rangle=0$
in the spherical-polar coordinate $(r,\theta,\phi)$
implies that the potential must be in the form of
$\Phi(r,\theta,\phi)=f(r)+r^{-2}g(\theta,\phi)$,
provided that $(\langle v_r^2\rangle-\langle v_\theta^2\rangle)
(\langle v_r^2\rangle-\langle v_\phi^2\rangle)
\ne\langle v_\theta v_\phi\rangle^2$.
\end{corol}
Since $(h_r,h_\theta,h_\phi)=(1,r,r\sin\theta)$,
we have $h_\phi/h_\theta=\sin\theta$, which is independent of $r$.
Theorem \ref{theo-ps} therefore indicates
\begin{equation}\begin{split}
\dst_{r\theta}(\Phi)
&=\frac{\partial^2\Phi}{\partial r\partial\theta}
+\frac2r\frac{\partial\Phi}{\partial\theta}
=\frac1{r^2}
\frac\partial{\partial\theta}\left[\frac{\partial(r^2\Phi)}{\partial r}\right]=0
;\\\dst_{r\phi}(\Phi)
&=\frac{\partial^2\Phi}{\partial r\partial\phi}
+\frac2r\frac{\partial\Phi}{\partial\phi}
=\frac1{r^2}
\frac\partial{\partial\phi}\left[\frac{\partial(r^2\Phi)}{\partial r}\right]=0,
\end{split}\end{equation}
and so follows that $\partial(r^2\Phi)/(\partial r)$
is a function of $r$ alone.
Consequently, the general solution for $\Phi$ is given by
$\Phi(r,\theta,\phi)=f(r)+r^{-2}g(\theta,\phi)$.

The same result is also obtained in any 3-dimensional coordinate
$(r,q^2,q^3)$ such that the coordinate surfaces of constant $r$
consist of the set of concentric spheres (of the radius $r$)
and $(q^2,q^3)$ corresponds to the coordinate on the unit sphere.
The notable example of such coordinates other than the spherical
coordinate is the conical coordinates \citep[see][]{MF53}.
The scale factors for such a coordinate system
are found to be $h_r=1$, $h_2=r\hat h_2(q^2,q^3)$ and $h_3=r\hat h_3(q^2,q^3)$,
for which $h_2/h_3=\hat h_2/\hat h_3$ is independent of $r$, and so
\begin{equation}
\dst_{ri}(f)=\frac{\partial^2f}{\partial r\partial q^i}
+\frac2r\frac{\partial f}{\partial q^i}
=\frac1{r^2}\frac{\partial^2(r^2f)}{\partial q^i\partial r}
\end{equation}
where $i\in\set{2,3}$.
Then $r^2\dst_{ri}(h_j^{-2}) =\partial^2(\hat h_j^{-2})/(\partial
q^i\partial r)=0$ for $i,j\in\set{2,3}$, whereas
$r^2\dst_{ri}(h_r^{-2})=0$. Again from equation
(\ref{eq:jm}), we then find that, if all second and fourth moments
with an odd power of $v_r$ vanish in such a coordinate,
$\partial(r^2\Phi)/(\partial r)$ is a function of $r$ alone and so
$\Phi=f(r)+r^{-2}g(q^2,q^3)$.

In general, any single-valued (smooth) function on the unit sphere
$g(\theta,\phi)$ may be expressed as the sum over the spherical harmonics
as in $g(\theta,\phi)=\sum_{\ell,m}c^\ell_mY_\ell^m(\theta,\phi)$, for which
\begin{equation}
\nabla^2\left(\frac g{r^2}\right)
=\frac1{r^4}\sum_{\ell=0}^\infty(1-\ell)(2+\ell)
\left(\sum_{m=-\ell}^\ell c^\ell_mY_\ell^m\right).
\end{equation}
Similar to the axisymmetric case, the $r^{-4}$-density singularity as
$r\to0$ is again unphysical as it is unintegrable. Provided that the
tracer population includes the orbits passing the center, physical
potentials consistent with $\partial(r^2\Phi)/(\partial r)$ being a
function of $r$ alone should thus be spherically symmetric. Although
the dipole potential (corresponding to $\ell=1$) is formally allowed,
there is no real source generating such potentials. In principle,
masses within a fixed boundary may be arranged in such a way that the
potential outside the boundary becomes dipole-like.
However, for such cases, the potential
within the boundary cannot be dipole-like without it possessing the
$r^{-4}$-singularity at the center. Thus, in order for it to avoid
the unintegrable singularity, the potential within the boundary should
no longer be separable in the spherical coordinate. This may still be
acceptable if all orbits in the tracer population are restricted to
the outside of the boundary \citep[see e.g.,][Appendix A]{Ev15}.

\section{Conclusions}

This paper has shown that the alignments of the velocity moments of a
stellar system can provide powerful constraints on the potential. This
idea may be traced back to the classical work of \citet{Ed15} and
\citet{Ch39}. Their papers however muddied the issue by assuming
unnecessarily restrictive forms for the distribution.
%
By contrast, \citet{Ev15} have shown that, if the velocity distribution
in a steady state possesses planes of reflection symmetry such that
$\df(-v_1,v_2,v_3;\bm r)=\df(v_1,v_2,v_3;\bm r)$
and similarly for $v_2$ and $v_3$, then
the potential must be of St\"ackel. We can recast this
result in terms of the velocity moments. Suppose all the mixed
second moments vanish
(i.e.\ $\langle v_iv_j\rangle=0$ for $i\ne j$) so that
the ``stress tensor'' is aligned in some St\"ackel coordinate system.
Then, if the only non-vanishing fourth moments are those in
the form of $\langle v_i^4\rangle$ or $\langle v_i^2v_j^2\rangle$,
then the potential must be separable in the same coordinate. Although our
conclusions are superficially very similar to those of \citet{Ed15},
our work is much more general in its scope, as nothing
has been assumed about the distribution other than
some basic symmetries.

Our work has been motivated by the stellar halo of the Galaxy, for
which the second moments do appear to be close to spherical alignment.
It is worth stating the form of our result explicitly in the
spherical-polar coordinate system.  If the second velocity moments are
spherically aligned and all the fourth velocity moments with the
radial component $v_r$ being either linear or cubic vanish,
the potential must be separable in the spherical coordinate.
An alternative way to state this is, if the second velocity moments
are spherically aligned and the velocity distribution is symmetric
with respect to $v_r$, then the potential must be
$\Phi=f(r)+r^{-2}g(\theta,\phi)$. Although current observational
  studies \citep[e.g.,][]{SEA,Bo10,Ki15} seem to indicate that the
  condition required to infer the Galactic potential to be separable
  in the spherical coordinate is present, a word of caution is
  still warranted before any definite conclusion on the shape of the
  Galactic halo is reached. Most of these data only cover a relatively
  small volume of the Galaxy and there is a large extrapolation to go
  from the local velocity ellipsoid being radially aligned (within
  observational uncertainties) to the global alignments of the
  velocity moments. However, with upcoming availability of large data
  sets including proper motions for many stars in substantial local
  volumes, it is within our grasp in near future to test
  symmetry properties of the ``global'' velocity distribution explicitly
  for the stellar halo \citep[c.f.][]{Ev15}.

Eddington's paper is now a hundred years old. It is a testimony to his
greatness that it remains a fruitful avenue for research still today.
Perhaps the most interesting topic for future exploration is to
understand how insights from the St\"ackel models with their exact
alignment can be applied to numerical models based on orbital tori or
made-to-measure \citep{BM11,Ev15}. In the framework of the
  Hamiltonian perturbation theory, the invariant orbital tori of
  regular orbits found in non-St\"ackel potentials may be understood
  as the result of perturbation to the exactly integrable St\"ackel
  models. In fact, some recent works \citep[e.g.,][]{Bi12,SB14,BRM}
  have used the St\"ackel potentials to find an approximate third
  integral of motion (which sometimes doubles as an action integral)
  to define regular orbits in more realistic (non-St\"ackel)
  potentials. It is then an interesting question how the conclusion of
  the present paper relates to the behavior of the distribution
  consisting of these regular orbits in non-St\"ackel potentials. It
  appears that each regular orbit in these potentials observes its own
  St\"ackel coordinate system, which seems to suggest that these
  systems are able to evade the global constraint required for the
  St\"ackel models, but any definite statement should follow more
  careful studies. In this regards, we hypothesize that the St\"ackel
  potentials (including those due to Noether--Killing symmetries) are
  the only potentials in which \emph{all} (bound) initial conditions
  result in a regular orbit, whereas, in other potentials, there must
  exist some initial conditions that leads to an irregular (which can
  be chaotic or ergodic) orbit.

\bigskip
\acknowledgments\noindent\small
Work by JA is supported by the CAS Fellowships
for Young International Scientists (Grant No.:2009Y2AJ7) and
grants from the NSFC. 


\end{CJK*}
\end{document}